# Mean Field MARL Based Bandwidth Negotiation Method for Massive Devices Spectrum Sharing

Tianhao Li[1,2], Yu Tian[2], Shuai Yuan[2], Naijin Liu[1,2]
[1] *School of Aerospace Science and Technology (SAST), Xidian University*
[2] *Qian Xuesen Laboratory of Space Technology, China Academy of Space Technology*

*Abstract*—In this paper, a novel bandwidth negotiation mechanism is proposed for massive devices wireless spectrum sharing, in which individual device locally negotiates bandwidth usage with neighbor devices and globally optimal spectrum utilization is achieved through distributed decision-making. Since only sparse feedback is needed, the proposed mechanism can greatly reduce the signaling overhead. In order to solve the distributed optimization problem when massive devices coexist, mean field multi-agent reinforcement learning (MF-MARL) based bandwidth decision algorithm is proposed, which allow device make globally optimal decision leveraging only neighborhood observation. In simulation, distributed bandwidth negotiation between 1000 devices is demonstrated and the spectrum utilization rate is above 95%. The proposed method is beneficial to reduce spectrum conflicts, increase spectrum utilization for massive devices spectrum sharing.

## I. INTRODUCTION

The emergence of worldwide industrial changes and the development of new fields and markets have placed additional demands and challenges on existing mobile communication technologies. As one of the new scenarios introduced by the fifth-generation cellular communication system (5G), massive wireless communication networks provide connectivity for large-scale heterogeneous wireless devices to support the development of new areas such as mMTC, Internet of Things (IoT), and smart factory. According to Ericsson, IoT connections will reach 26.9 billion by 2026, of which cellular IoT connections will be close to 6 billion [1]. In such large-scale scenario, spectrum resource management becomes particularly important when so many various devices simultaneously access the spectrum in the same area.

Traditional wireless multiple access technologies, which can be classified as contention based and schedule based [2], will break down in massive networks. Typical examples of contention-based schemes are ALOHA and Carrier Sense Multiple Access (CSMA). This type of method has high efficiency in scenarios with sparse number of devices, but when the number becomes enormous, the access collision becomes critical, making the transmission efficiency drop significantly. For schedule-based schemes such as TDMA, FDMA, CDMA, etc., they allow conflict-free access to the shared spectrum. However, when devices are massive, heterogeneous and dynamic, it can cause serious communication overhead and computational burden, and has high requirements on the efficiency and performance of the central AP. As a result, massive networks will pose two major challenges: increased probability of conflict for spectrum access and complex resource management [3]. In order to achieve efficient spectrum resource management for massive wireless networks, a new system architecture must be constructed and a corresponding spectrum sharing method must be proposed.

This paper proposes a negotiation-based system architecture to solve the problem of spectrum sharing for large-scale devices and applies the mean field multi-agent reinforcement learning (MF-MARL) algorithm to enable each device to learn local negotiation policies. Mean field multi-agent reinforcement learning was developed in [4] for solving the reinforcement learning (RL) problem when a large number of agents coexist, and the convergence was discussed under certain assumptions. Mean field (MF) theory [5] approximates all the influences on a single agent in a stochastic process model as a single external field, thus decomposing the multi-agent problem into multiple single-agent problems to be solved. Therefore, the application of MF to massive spectrum sharing problem has a great potential, as it simplifies the interactions between agents and is well suited for system scalability. This paper proposes a spectrum sharing method based on MF-MARL, in which each agent abstracts a virtual agent by observing the bandwidth requests of a subset of agents which is called neighbors, to effectively overcome the instability of a multi-agent environment. Instead of learning directly with all other agents, each agent learns from the interactions with its virtual agent and eventually converge to the optimal policy. Individual learning is based on the dynamics of the group, while the group's behavior is updated according to the individuals. Finally, according to the problems that appeared during implementation, two techniques are proposed to stabilize the training and convergence process.

The contributions of this paper are summarized as follows:

- A bandwidth negotiation mechanism for spectrum sharing of massive wireless devices is established, in which devices send bandwidth requests to AP based on local information exchange, and the overall performance is evaluated by AP.
- Based on MF-MARL algorithm, the distributed spectrum negotiation is modeled as an incomplete stochastic game / Markov game with imperfect information, and the negotiation strategy of each device is obtained using the distributed learning algorithm with centralized feedback.
- Two practical techniques are proposed to solve the overestimation bias problem of the training process and stabilize the update of the average actions during convergence. From the simulation results, it can be found that the techniques significantly increase the stability and performance of the algorithm.

The rest of the paper is organized as follows. Sec. II provides a review of the referenced MARL algorithms. Sec. III explains the bandwidth negotiation mechanism designed for the spectrum sharing of large-scale wireless communication networks. Sec. IV shows the proposed distributed negotiation policies learning method and two practical technology spectrum sharing methods. Sec. V presents and analyzes the simulation results. At last, the conclusion is drawn in Sec. VI.

## II. MARL PRELIMINARIES

This section provides a preliminary introduction to MARL theory and presents two algorithms applied in this paper: one is independent deep Q learning (IDQL) which directly applies single-agent RL to multi-agent problems, and the other is mean field MARL proposed in [4] which is the underlying technique of the spectrum sharing method proposed in this paper to solve the problems arising in scenarios with large-scale agents.

MARL is a combination of reinforcement learning and game theory, which consists of an environment and multiple agents gaming each other. Agents make local decisions and impose joint action on the environment, while the environment feedback state transfers and rewards to the agents. Eventually, the learning algorithm allows each agent's policy to converge and reach a Nash equilibrium of the system.

### A. Independent Deep Q Learning

Each agent learns a local policy $\pi^j$ independently and treats the behavior of other agents as part of the environment [6]. The algorithm approximates the local Q function, i.e., the action-value function, by training a DQN, which evaluates the value of taking a local action $a^j$ in a certain state $s$. For an independent agent, the definition of the Q function can be formulated as:

$$Q^j(s, a^j) \triangleq \mathbb{E}\left[\sum_{t=0}^{\infty} \gamma^t r_t^j \,|\, s, a^j, \pi^j\right] \quad (1)$$

where $r_t^j$ indicates the reward from environmental feedback at time step $t$, and $\gamma \in (0,1)$ denotes the reward discount factor. The training of the DQN is done by minimizing the network loss through gradient descent algorithm, and usually "experience replay" and "target neural network" are also applied to stabilize the training process. Eventually, the object of IDQL is to obtain a locally optimal policy $\pi_*^j$ such that $Q_*^j = \max_{\pi^j} Q^j(s, a^j)$.

Due to the dependency on theoretical convergence guarantees in Markov decision process, IDQL has good performance in single-agent scenarios. However, when it comes to multi-agent scenarios, each independent agent directly integrates others into the environment, while at the same time its optimal policy is affected by the changes in the other agents' policies, which leads to a dynamic environment problem that prevents all agents from learning the optimal local policies.

### B. Mean Field MARL

In order to deal with the RL problems in large-scale agent scenarios, mean field MARL algorithm is proposed, which not only solves the dynamic environment problem, but also simplifies the interactions between agents and guarantees the convergence of the algorithm under certain assumptions.

First of all, the dynamic environment can be handled by considering the actions of all agents, not just local actions, when estimating Q functions that can be defined as:

$$Q^j(s, \boldsymbol{a}) \triangleq \mathbb{E}\left[\sum_{t=0}^{\infty} \gamma^t r_t^j \mid s, \boldsymbol{a}, \boldsymbol{\pi}\right] \quad (2)$$

where $\boldsymbol{a} \triangleq [a^0, a^1, \ldots, a^{N-1}]$ denotes the joint action of all agents, and $\boldsymbol{\pi} \triangleq [\pi^0, \pi^1, \ldots, \pi^{N-1}]$ is the joint policy. Nevertheless, since the dimension of joint action increases with the number of agents, it is difficult for agents to learn the Q function when the scale is extremely large. Therefore, a simplified approach is to decompose the Q function into an average of the Q values with respect to multiple pairs of agent $j$ and its neighbors $N(j)$, and further abstract all neighbors into a virtual agent using the idea of MF theory. The expression is shown below:

$$Q^j(s, \boldsymbol{a}) = \mathbb{E}_{k \in N(j)}\{Q^j(s, a^j, a^k)\} \approx Q^j(s, a^j, \bar{a}^j) \quad (3)$$

where $\bar{a}^j = \mathbb{E}_{k \in N(j)}\{a^k\}$ is defined as the average action of agent $j$'s neighbors, i.e., the action of its virtual agent. A detailed derivation of the approximation of the above equation based on Taylor polynomial can be found in [4].

Using the MF approximation, the MARL problem is transformed into solving for the best response of agent j to the average action of its neighbors $\bar{a}^j$, leading to a direct dependence of the local policy $\pi^j$ on the observation of the neighbors' actions. Exploration and exploitation are also the key point to decide whether the RL algorithm can reach the optimal solution. For trade-off, a gradual annealing ε-greedy policy is used in this paper:

$$\pi^j(a^j | s, \bar{a}^j) \leftarrow \begin{cases} 1 - \varepsilon + \frac{\varepsilon}{|A^j|}, & \text{if } a^j = \underset{a^j}{\arg\max}\, Q^j(s, a^j, \bar{a}^j) \\ \varepsilon / |A^j|, & \text{else} \end{cases} \quad (4)$$

## III. SPECTRUM NEGOTIATION MECHANISM

The considered spectrum sharing system is shown in Figure 1, which consists of an access point (AP), $N$ active wireless devices, $C$ frequency division multiplexing data channels, a control channel. The proposed spectrum negotiation mechanism works as follow. Firstly, all active devices exchange their bandwidth requests with their neighbor devices (in frequency domain) using D2D communication. Secondly, each device updates its bandwidth request based on its own demand and the received messages from neighbors. The first and second steps are repeated for certain periods. Then all devices send their negotiated request to AP over the control channel. AP broadcast an evaluation of global performance to each device through the control channel.

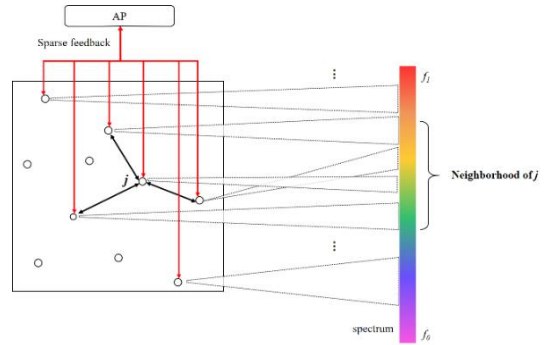

Figure 1. Proposed spectrum negotiation mechanism

The goal of spectrum negotiation is to make the best use of scarce spectrum resources while ensuring fairness for all heterogeneous devices, which means the total bandwidth usage should approaches the bandwidth of available spectrum as close as possible. If the sum of requested channels is less than the total number of data channels, there are idle channels resulting in low spectrum utilization; while if the sum is greater than the total number, some agents are not accessible and the fairness of the sharing is poor.

In summary, the proposed spectrum negotiation mechanism has several advantages over traditional centralized or distributed systems, especially in wireless systems with large-scale coexistence of heterogeneous devices, as follows.

(1) In a centralized system, the AP directly coordinates the bandwidth of all devices, a process that requires frequent information interaction between AP and each device, as shown in Figure 2. This not only requires AP to have extremely high computational efficiency, but also increases the signaling overhead of the control channel. Instead, the proposed spectrum negotiation mechanism decentralizes the decision-making to each device, which releases the AP processing bottleneck while increasing the scalability of the system. In addition, by introducing D2D-based negotiation between devices and a delayed feedback mechanism, the frequency of message interaction between devices and AP is reduced and the bandwidth bottleneck of the control channel is minimized.

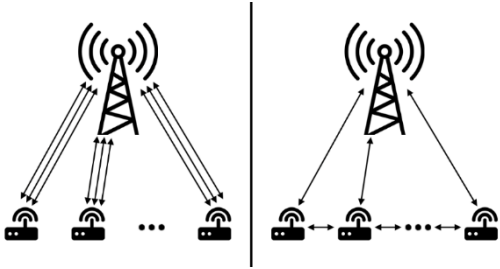

Figure 2. (a) Centralized negotiation scheme. (b) the proposed negotiation scheme.

(2) Compared to fully distributed systems with only local information, AP provides global feedback for devices' decision, which is beneficial for improving the stability and convergence of the distributed algorithm. Each device shares its bandwidth request information only to its neighbors instead of all devices in the network, minimizing the signaling overhead of the system while maintaining system effectiveness.

The core of proposed spectrum negotiation method lies in designing an efficient distributed decision algorithm. This algorithm should convert local observations into reasonable bandwidth usage requests, the joint of which achieves the globally optimal spectrum sharing. To solve the distributed decision problem, a spectrum negotiation algorithm is proposed based on the MF-MARL, which provides outstanding performance especially in scenarios where massive devices coexist.

## IV. REINFORCEMENT LEARNING SOLUTION

In this section, the MF-MARL algorithm is applied to learn and optimize the local negotiation strategies in the above spectrum sharing system. Firstly, the MF-MARL algorithm is adapted to the spectrum negotiation process which is constructed as a game model. In addition, two practical techniques are proposed in order to stabilize the training process and convergence process of the algorithm: clipped double DQN and soft update of average actions. Finally, a complete summary of the proposed spectrum sharing algorithm is presented.

### A. MF-MARL Based Spectrum Negotiation

The spectrum negotiation with distributed decision making can be defined as a partially observable stochastic game / Markov game $\Gamma \triangleq (S, A^0, \ldots, A^{N-1}, r^0, \ldots, r^{N-1}, \gamma)$, where $S$ is the global state, $A^j$ is the action space of agent $j$, $r^j$ is the reward obtained by $j$, and $\gamma$ is the reward discount factor. To be specific, the state space, action space, and reward function are defined as follows:

**State space:** The global state is defined as the set of currently active devices $\mathcal{N} = \{0, 1, \ldots, N-1\}$, but it is impractical to obtain for devices due to the huge quantity. Therefore, each device $j$ only acquires its neighbor's information $N(j) \subset \mathcal{N}$ as local observations.

**Action space:** The frequency division multiplexed data channel contains a series of subchannels $\mathcal{C} = \{0, 1, \ldots, C-1\}$, and the bandwidth of device is expressed as the utilized number of subchannels. Each device $j$ takes an action $a^j \in A^j$ that indicate the requested bandwidth of data channel sent to AP. The optional number of subchannels $A^j$ varies with the traffic type of device $j$, and is limited by the minimum acceptable quality and maximum demand quality of connections.

**Reward function:** The reward function drives the learning process in MARL, and decisions are made by maximizing the expected rewards. With the objective of maximizing channel utilization while ensuring that the sum of the requested bandwidth does not exceed the total bandwidth of the data channel, the reward function is defined as:

$$r^0 = \cdots = r^{N-1} = r$$

$$= \begin{cases} -\left(1 - \sum_{j \in \mathcal{N}} a^j / C\right), & if \sum_{j \in \mathcal{N}} a^j \leq C \\ -1, & else \end{cases} \quad (5)$$

When the sum of the requested bandwidth is less than the total number of subchannels, it means that the data channel is not fully utilized and the reward decreases as the number of free subchannels increases; and when the requests exceed the total number, it means that there are devices whose requests cannot be fulfilled so all devices obtain the lowest reward.

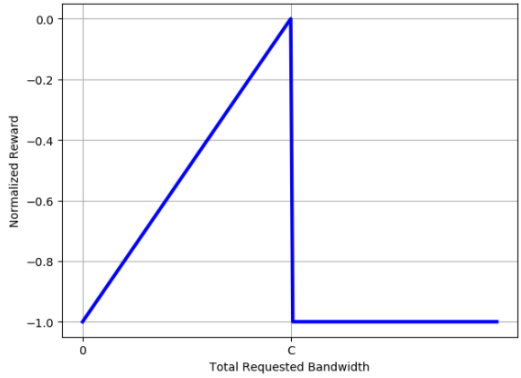

Figure 3. Global reward function with total requested bandwidth

During spectrum negotiation, each device $j$ does not know the global state $S$, but it is able to obtain and react to the actions of its neighbors $\{a^k\}, k \in N(j)$ and its reward $r^j$ from AP. Each agent $j$ learns a local policy $\pi^j$ that determines its requested bandwidth, and all agents form a joint policy $\boldsymbol{\pi} \triangleq [\pi^0, \dots, \pi^{N-1}]$. Based on above defined stochastic game model, the MF-MARL algorithm mentioned in Sec. II can be applied, by which each device finally gets an effective spectrum negotiation strategy.

### B. Implementation with Practical Techniques

The MF-MARL algorithm is realized by approximating the Q function using a deep neural network (DQN) with parameters $\phi^j$. For the purpose of solving the overestimation bias problem during the training process and the oscillation phenomenon of the average action during the convergence process, two practical techniques are proposed and analyzed in detail as follows.

*1) Clipped Double DQN:* In deep Q-learning with discrete actions, errors caused by function approximation are unavoidable. Updating the Q value estimations with error-existing greedy targets will typically have values greater than the true values [7]. Furthermore, propagation through the Bellman equation leads to overestimation bias in the Q values and makes errors in the policy generated.

Although Double DQN was proposed in [8] to address this problem, as the networks slowly change, the current and target networks converge, making the bias severe over training. [9] proposed Clipped Double Q-Learning to solve the problem under the Actor-Critic architecture, which is based on the simultaneous estimation of two equivalent networks and taking the smaller of the two values. Based on that, a DQN version of the Clipped Double Q-Learning algorithm is proposed here, which contains a pair of independent estimation networks $(Q_{\phi_1}, Q_{\phi_2})$ as well as a pair of target networks $(Q_{\phi'_1}, Q_{\phi'_2})$. With the off-policy learning, the loss function can be defined as:

$$L(\phi_1^j, \phi_2^j) = \sum_{i=1,2} \left(y^j - Q_{\phi_i^j}(s, a^j, \bar{a}^j)\right)^2 \quad (6)$$

where the target value is considered as the smaller value of the greedy strategy outputs of the two target networks:

$$y^j = r^j + \gamma \min_{i=1,2} Q_{\phi_i'^j}(s', a^j, \bar{a}^j) \quad (7)$$

*2) Soft Update of Average Actions:* During implementation of the algorithm, a problem was noticed in that there was a high chance that the algorithm would not stabilize to a spectral negotiation result, as demonstrated by the fact that most agents repeatedly changed the requested bandwidth during the post-training period. This phenomenon arises because the negotiation policies of devices learning relies on observations of neighbors, which makes it sensitive to other agents' behaviors, and the accumulation of small changes in individual behavior can lead to large changes in the group behavior. The constant interaction between individuals and groups makes this problem particularly severe.

Instead of directly observing neighbor actions, a soft update technique is proposed for tracking the average

action of neighbors $\bar{a}^j$, which is formulated as:
$$\bar{a}^j_t \leftarrow \alpha \bar{a}^j_{t-1} + (1-\alpha)\bar{a}^j \quad (8)$$
where $\bar{a}^j$ is the currently observed average action of the neighbors, and $\bar{a}^j_t$ is a temporal iteration of the average actions used to weaken the sensitivity to the neighbor actions. The smoothing factor $\alpha \in [0,1)$ is defined as the weight of the previous average action retained by each update and directly influences the sensitivity of the system. When $\alpha = 0$, the algorithm turns back to hard update and the oscillation problem becomes critical at the end of training; however, when $\alpha = 1$, the algorithm transforms to IDQL which does not obtain information from other agents, and the problem of dynamic environment occurs without effective training. Therefore, the factor $\alpha$ needs to be carefully designed to ensure the best convergence, which will be discussed in Sec. V.

### C. Complete Algorithm

In this paper, a spectrum sharing method for massive wireless networks based on mean field multi-agent reinforcement learning is proposed, and two practical techniques are further developed to optimize the algorithm performance. The entire algorithm procedure is presented as follows:

1) For each agent with initial connection to the wireless network, assign device index $j$, neighbor information $N(j)$ and total number of available data channels $C$. Initialize DQN parameters $\left(Q_{\phi^j_i}, Q_{\phi'^j_i}\right)_{i=1,2}$, average action $\bar{a}^j_0$ and smoothing factor $\alpha$.

2) During the spectrum negotiation, agent $j$ makes a bandwidth request $a^j$ to AP according to the local policy $\pi^j$ and share it with its neighbors. After receiving the requests of neighbors $\{a^k | k \in N(j)\}$, soft update of average action $\bar{a}^j_t$ is performed.

3) After each iteration, agent $j$ receives a global reward $r$ from AP based on the overall status of all requests, and then stores the experience $<N(j), a^j, r, \bar{a}^j_t>$ into memory buffer.

4) Using the training technique as above, sample a batch of experiences from buffer $\mathcal{B}$ to train the agents' DQN. Repeat the spectrum negotiation process from 2) to 4) until local policies converge.

5) After negotiation, AP divides the data channel according to the last bandwidth requests and inform all devices of their allocated subchannels.

The above procedure can be concluded as Algorithm 1.

| **Algorithm 1** MF-MARL based Spectrum Sharing |
|---|
| 1: Initialize evaluation networks with parameters $\phi^j_1, \phi^j_2$ |
| 2: Initialize target networks $\phi'^j_1 \leftarrow \phi^j_1, \phi'^j_2 \leftarrow \phi^j_2$ |
| 3: Initialize action of virtual agent $\bar{a}^j_0$ and buffer $\mathcal{B}^j$ |
| 4: **while** not converged **do** |
| 5:    **for** $j = 0,1,\dots,N-1$ **do** |
| 6:       Sample $a^j$ based on $Q_{\phi^j_1}$ by Eq. (4) with $\bar{a}^j_{t-1}$ |
| 7:       Obtain new $\bar{a}^j$, and compute $\bar{a}^j_t$ by Eq. (8) |
| 8:       Take joint action $\mathbf{a}$ and receive $r$ from AP |
| 9:       Save $<N(j), a^j, r, \bar{a}^j_t>$ into the buffer $\mathcal{B}^j$ |
| 10:    **for** $j = 0,1,\dots,N-1$ **do** |
| 11:       Sample a batch of experiences from $\mathcal{B}^j$ |
| 12:       Calculate training target $y^j$ by Eq. (7) |
| 13:       Update networks by minimizing loss $L(\phi^j_1, \phi^j_2)$ |
| 14:       Update the target networks with learning rate $\tau$: |
| 15:          $\phi'^j_i \leftarrow \tau \phi^j_i + (1-\tau)\phi'^j_i$ |

## V. PERFORMANCE EVALUATION

In this section, simulation results are shown to illustrate the performance of the proposed algorithm in comparison to IDQL approach which has no information interaction between independent devices. Training of the spectrum negotiation strategy for each device is performed in a distributed manner, with each DQN consisting of two fully connected hidden layers of 64 units each. At the beginning, the greedy coefficient $\varepsilon$ is set to 0.9 and then slowly reduced to 0 as the training progresses. Each DQN is trained for 5,000 iterations with a learning rate of 0.001, and after that, the final spectrum sharing and the achieved performance by the system are recorded. The main parameters are given in Table 1.

Table 1. Simulation Parameters

| Parameter | Value |
|---|---|
| $N$ | 60, 100, 300, 500, 1000 |
| $|N(j)|$ | $0, 1, 2, \dots, N-1$ |
| $C$ | 500 |
| $A^j$ | $A^j \subseteq \{0, 1, \dots, 9\}$ |
| $\alpha$ | 0.0, 0.2, 0.5, 0.7, 0.9, 0.99 |
| $|\mathcal{B}|$ | 1000 |

To compare the convergence of the algorithm for scenarios with different scales, Figure 4 shows the loss of the training process when the number of devices varies from 60, 100, 300, 500 to 1000. It can be found that the number of steps required for the training process is positively related to the device scale, while the growth of cycles tends to slow down as the quantity increases. In contrast, the IDQL algorithm maintains a high loss and its training does not converge well to the optimal policy, and the reason is that the dynamic environment problem makes it impossible for the IDQL algorithm to explore efficiently, which prevents convergence to the optimal policy.

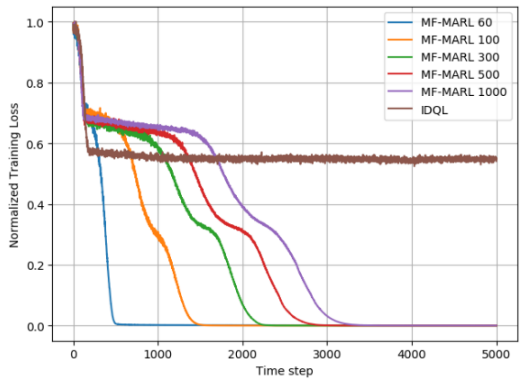

Figure 4. Normalized training loss over time steps for different number of devices

Figure 5 indicates the channel utilization rate after algorithm convergence for different number of devices. The better the devices perform, the higher the percentage of utilized channels, which means that spectrum resources are fully utilized, while every device has the accessing opportunity. From the figure, it can be found that the proposed approach converges well to the optimal channel usage in scenarios with any number of devices. For comparison, the IDQL approach performs well in scenarios with small-scale devices, while the performance decreases dramatically when the number of devices expands, and has a significantly low channel utilization in large-scale scenarios. Meanwhile, the results in Figure 6 show the average number of channels utilized by each device after the algorithm converges. MF-MARL is very close to the optimal value, but the IDQL approach has a significantly lower average action in large-scale scenarios. One possible reason is that devices cannot obtain optimal policies based on the dynamic environment, instead conservative policies are adopted which reduce the occupied resources to prevent excessive overall channel requests.

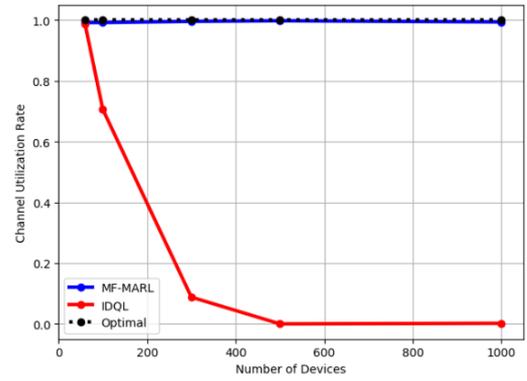

Figure 5. Variation of channel utilization with number of devices

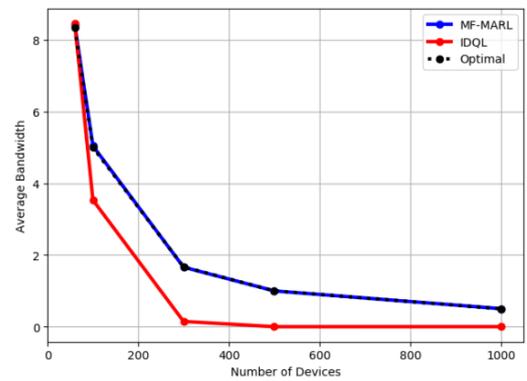

Figure 6. Variation of average requested bandwidth with number of devices

In order to investigate the impact of change in number of neighbors on the algorithm, Figure 7 shows the variation in channel utilization rate corresponding to the number of neighbors from 0 to 299 when the system includes 300 devices. It can be found that the higher the number of neighbors, the higher the achieved channel utilization. When the quantity is too low, the average action of neighbors has great instability and the dynamic environment problem is raised, making it difficult for devices to obtain effective training. When neighbors are set to 0, the algorithm is equivalent to IDQL with very low channel utilization.

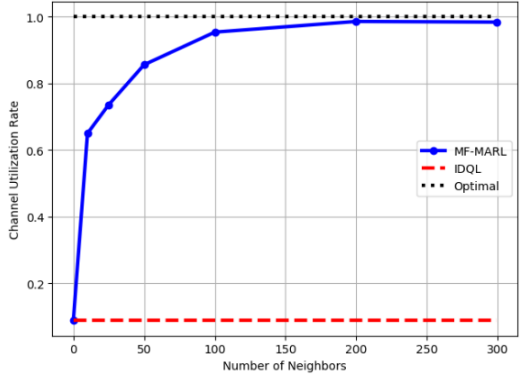

Figure 7. Variation of channel utilization with number of neighbors

During simulation, a phenomenon was found that after the training loss became extremely small, devices become very sensitive to their neighbors' action, making the system unable to converge stably. Therefore, a soft update of average actions is proposed and the effect of the smoothing factor $\alpha$ on the convergence of system is investigated. Figure 8 shows that the smoothing factor contributes significantly to the stability of the system, and larger factors significantly attenuate the oscillation of the devices' average action. As illustrated in Figure 9, besides improving stability, the technique enhances the system's utilization of spectrum resources and outperforms the IDQL algorithm for all values. With more stable observations, each device slowly changes its requests, allowing the negotiation process more stable and making the algorithm potentially converge to better performance.

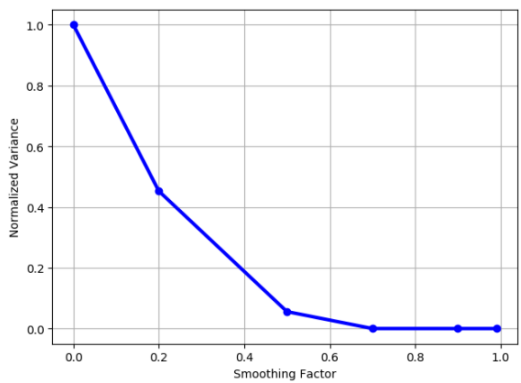

Figure 8. Normalized variance of average requested bandwidth with different smoothing factors

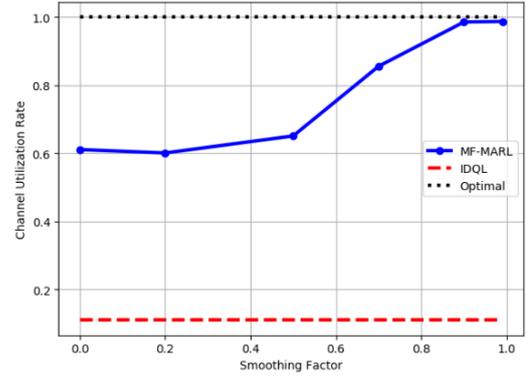

Figure 9. Average channel utilization with different smoothing factors

## VI. CONCLUSION

This paper proposes a bandwidth negotiation-based system architecture for large-scale wireless devices spectrum sharing, and a distributed negotiation policy learning method based on the MF-MARL algorithm. The simulation results show that the MF-MARL method can solve the dynamic environment problem well, and achieve a higher channel utilization rate in the case of large-scale devices. Similarly, the number of neighbors observed by each device is also an important factor affecting the performance of the algorithm. A large number of neighbors makes the training process more stable and efficient, but there is a trade-off between channel utilization and energy consumption. Finally, with the help of two practical techniques, the negotiation process has better stability and higher channel utilization.